\documentclass[11pt,twoside]{article}

\newcommand{\f}{\varphi}
\newcommand{\A}{\cal A}
\newcommand{\hh}{(r_+-r_-)}
\newcommand{\M}{\tilde{M}}

\begin{document}

\vspace*{1.cm}

\begin{center}
{\large\bf  On Black Hole Horizon Fluctuations }\\

\vspace*{.5cm}

K.L.~Tuchin\footnote{e-mail:tuchin@post.tau.ac.il}\\

\vspace*{.5 cm}

{\it Raymond and Beverly Sackler Faculty of Exact Science,}\\
{\it School of Physics and Astronomy,}\\
{\it Tel-Aviv University, Ramat Aviv, 69978, Israel}\\
 
\vspace*{.5 cm}

\today 
\end{center}

\vspace*{.3cm}
 
\begin{abstract} 
\vspace*{-.1cm}

 A study of the high angular momentum particles 'atmosphere' near the 
Schwarzschild black hole horizon  suggested that strong gravitational
interactions occur at invariant distance of the order of  
$\sqrt[3]{M}$. We present a generalization of this result to 
 the Kerr-Newman black hole case. It is shown that the 
larger charge and  angular momentum black hole bears, the
larger invariant distance at which strong gravitational interactions occur
becomes. This invariant distance is of order
$\sqrt[3]{r_+^2/\hh}.$
This implies, that the Planckian structure  of the Hawking radiation of
extreme black holes is completely broken.
\end{abstract}
\vspace*{.2cm}

PACS: 04.60.-m; 04.70Dy

{\em Keywords:  black hole, atmosphere, fluctuations, semi-classical
approximation}

\vspace*{.2cm}


\section{Introduction}\label{sec:introduction}

The black hole radiation is a direct consequence of non-static character
of the collapsing body metric. A wave propagated through the collapsing
body becomes red-shifted, hence an in-vacuum state defined with respect to
modes at past null infinity differs from an out-vacuum state, defined with
respect to modes at future null infinity. The red-shift is exponentially
large if collapse leads to the black hole formation.

It was shown by Hawking\cite{Hawking} that the black hole radiation
spectrum is the Planckian one, provided that there is no back reaction of
emerging quanta on metric; the matter fields propagate on a classical
background metric, and the wave equation is valid at all scales. 

This semi-classical approach of Hawking becomes doubtful at scales of the
order of $l_p=\sqrt{G\hbar /c^3}$, i.e. the ultraviolet part $\omega
>l_p^{-1}$ of the Hawking radiation spectrum is possibly non-Planckian.
However, such huge energies are of little interest from an experimentalist
point of view. It turns out that problems begin  at much smaller
scale. 
 
We can gain more by putting in question the possibility to  neglect  the
radiation back reaction. Existence of energy flux from the black hole at
infinity implies that the black hole mass decreases. However, as long  as 
the mass of the black hole is large compared to the Planck mass, the rate
of evolution of the black hole is small compared to the characteristic
time for the light to cross the gravitational radius\cite{Birrel}. Thus,
we can describe the black hole by a sequence of stationary solutions; in
each solution the back reaction influence on the radiation spectrum can be
neglected\cite{Hawking}.

On the other hand, despite the fact that the Riemann tensor is
small near the black hole horizon ${\bf R}\sim1/M^2$ the normal modes get
exponentially red-shifted\cite{+}. This implies, that strong  
gravitational interactions occur there and, possibly, alter the Planckian
character of the spectrum.

The brick wall model proposed by 't Hooft is a simple model that provides
some insight into the problem\cite{brick}. The idea is to calculate the
number of the (scalar) wave equation solutions near the horizon of the
Schwarzschild black hole. This number can be interpreted as a partition
function of  the black hole  "atmosphere" --- a gas of high angular
momentum Hawking particles reflected back by the space-time
curvature\cite{sanny}. The  atmosphere's entropy turns out to be divergent
at the horizon. However, one can introduce an invariant cut-off $\rho$ of
the order of 1 and independent of $M$, such that the  atmosphere's entropy
becomes exactly the Bekenstein-Hawking one. This implies that the strong
gravitational interactions occur at invariant distance of the order of 1.
Several other authors arrive at this conclusion using other
models\cite{+}.

However, some authors arrive at  opinion that $\rho$ is greater that
1\cite{+}. In particular, Casher {\it et. al.} showed that $\rho$ is of
the order $M^{1/3}$, provided that one takes account of the atmosphere's
thermal fluctuations. This implies that the Hawking radiation Planckian
structure is broken at $\omega >M^{-1/3}$. Therefore, unlike the Planckian
spectrum quanta, the real black hole radiation quanta are  correlated,
and information about the state of in-falling matter gets encoded out of
the horizon\cite{sanny}.

The brick wall model may be criticized for it deals with non-renormalized
stress tensor. We can avoid this difficulty as discussed in
ref.~\cite{sanny}. Both approaches yield the same result for the
atmosphere's entropy. So, we shall apply the brick wall model again in
order to obtain an expression for the charged and rotating black hole
entropy (sec.~\ref{sec:thermo} and \ref{sec:Kerr})\cite{Korea}.

 Along the way we shall learn more about the atmosphere. In particular, it
will be shown that the main contribution to the
partition function comes from the particles emitted with a charge and
angular momentum of the same sign as the black hole ones.

In section~\ref{sec:fluctuations} the atmosphere thermal fluctuations will
be studied. It will be shown that  strong gravitational interactions
near the charge black hole horizon occur  at an invariant distance of the
order of $[r_+^2/\hh ]^{1/3}$. In section~\ref{sec:Kerr} it will be
argued that this result is valid for the Kerr black hole also.

In section~\ref{sec:shock} we shall confirm result of
section~\ref{sec:fluctuations}
 by applying the shock wave model\cite{Hooft} to study the
gravitational interactions between in-falling and out-going particles. 

We discuss the results in section~\ref{sec:discussion}.   

\section{Thermal Properties of Atmosphere}\label{sec:thermo}


\subsection{ Charged Black Holes}\label{sec:charge}
The charged black hole metric (Reissner-Nordstr{\o}m geometry) is given
by:
\begin{equation}\label{metricC}
ds^2=\frac{D}{r^2}dt^2-\frac{r^2}{D}dr^2-r^2(d\theta ^2+\sin ^2\theta d\f
^2),
\end{equation}
where $D=r^2-2Mr+Q^2$, $M$ and $Q$ are the mass and the charge of 
black hole respectively. The event horizon is located at surface $r_+$
defined as 
\begin{equation}\label{hor}
r_{^+_-}=M^+_-\sqrt{M^2-Q^2}.
\end{equation} 
A distant inertial observer observes the Hawking radiation ---  energy
flux from the black hole. This flux has a Planckian structure  provided
that we neglect 
the back reaction of this radiation on the metric\cite{Hawking}. 
Therefore,  the Hawking radiation can be considered as a  black body
radiation in a thermal bath with the temperature $T_H$
\begin{equation}\label{temperatureC}
T_H=\frac{1}{8\pi M}(1-\frac{Q^4}{r_+^4})=\frac{r_+-r_-}{{\A}},
\end{equation}
where ${\A}=4\pi r_+^2$ is the horizon area\cite{Birrel}.

Consider a massless  neutral scalar field $\phi$ in the charged
black hole background. One can separate variables in the wave equation
$$
g^{\mu\nu}\nabla _{\mu}\nabla _{\nu}\phi =0
$$ 
and obtain the following expressions for the basis functions\cite{deWitt}:
$$
u_{Elm}(x)=r^{-1}R_{l}(r)Y_{lm}(\theta ,\f )e^{-iEt},
$$
where $l=0,1,2\ldots$, $m=-l,-l+1,\ldots ,l-1,l$, $Y_{lm}(\theta ,\f)$ are
the spherical harmonics, $E$ is the hamiltonian eigenvalue.    
The radial part $R_{l}(r)$ satisfies the following equation:
\begin{equation}\label{radial}
[\frac{d^2}{dr^{*2}}-V_{El}(r)]R_{l}(r)=0,
\end{equation}
where $r^*$ and $V_{El}(r)$ are defined by the following
relations:
\begin{equation}\label{tort.}
\frac{dr^*}{dr}=\frac{r^2}{D},
\end{equation}
and
\begin{equation}\label{V}
V_{El}(r)= -E^2+l^2\frac{D}{r^4}+\frac{2MrD}{r^6}.
\end{equation}   

 The centrifugal barrier is 
$$ 
V_{l}^{c}(r)=l^2\frac{D}{r^4}.
$$
It is attractive for $r_+\le r<r_1$ and repulsive for $r>r_1$, where 
$r_1$ is the  root of the equation $\frac{dV^c}{dr}=0$:
\begin{equation}
r_1=\frac{3}{2}\left( M+\sqrt{M^2-\frac{8}{9}Q^2}\; \right) , 
\end{equation}
The tunneling through the angular momentum barrier may be neglected for
all but the lowest angular momentum modes since $V_{l}^{c}\sim l^2$\cite{+}. 

We see that near the horizon there exists an  atmosphere of the neutral 
high angular momentum particles. In general, the same is also true for the
{\em charged} scalar particles which are defined with respect to
basis functions of  the following equation:
\begin{equation}\label{waveequation}
g^{\mu\nu}(\nabla _{\mu}-ieA_{\mu})(\nabla _{\nu}-ieA_{\nu})\phi .
\end{equation}
Let us study properties of this thermodynamical system. At
first we shall calculate the number of the wave equation solutions
$\Gamma $. As long as $M\gg 1$ we can rely on  WKB approximation. So we
get:
\begin{equation}\label{Gam}
\Gamma =(2\pi)^{-3}\sum_{g}\int dx^k dp_k;\;\; k=r,\theta ,\f ,
\end{equation}
where the sum runs over all degenerate energy levels.

The covariant radial momentum $p_r$ associated with the differential
operator $\hat{p}_r$ in the wave equation is given by\cite{Landau,Misner}:
\begin{equation}\label{prq}
p_r^2=D^{-2}(r^2E-eQr)^2-D^{-1}l^2.
\end{equation}
Here $e$ is the charge of the scalar particles.
Since the metric is spherically symmetric integration over  the $\theta$
degree of freedom in eq.~(\ref{Gam}) is trivial. Thus, 
$$
\Gamma(E)=\sum_l (2\pi )^{-2}\sum_m\int d\f\int dr\int dp_r\approx
\frac{1}{2\pi}\int dl(2l+1)\int dr\int dE\frac{dp_r}{dE}.
$$
This expression is divergent at the horizon and at infinity. To proceed
further we should introduce  cut-offs. To this end let us use the brick
wall model \cite{brick}.

We introduce the following boundary conditions: $\phi (x)=0$ if $r\ge L$,
and $\phi (x)=0$ if $r\le r_++h$, where $L$ and $h$ are the infrared and
ultraviolet cutoffs respectively. We are interested in the contribution of
the horizon, i.e. in the term $O(h^{-1})$.
The term proportional to $L^3$ is the usual contribution from the vacuum
surrounding the system at large distances and is of little relevance here.
So we find: 
\begin{equation}\label{D}
D\approx \hh h,
\end{equation}
$$
\Gamma(E)\approx\frac{1}{\pi}\int dE
\int_0^{l_m}dl\, l\frac{r_+^4(E-e\Phi )D^{-2}h}
                        {\sqrt{D^{-2}r_+^4(E-e\Phi )^2-D^{-1}l^2}}
$$
\begin{equation}
=\frac{1}{\pi h}\frac{r_+^6}{\hh ^2}\int dE(E-e\Phi )^2,
\end{equation}
where $l_m$ is the largest possible value of $l$ which arises from the
obvious condition $p_r^2\ge 0$:
$$
l^2\le\frac{r_+^4(E-e\Phi)^2}{h\hh ^2},
$$
and $\Phi =Q/r_+$ is the electric potential at the horizon.

The free energy of the atmosphere is given by:
\begin{equation}
F = T_H\int d\Gamma (E)\ln(1-e^{(e\Phi -E)\beta}).
\end{equation}
This integral exists only if the condition $e\Phi -E\le 0$
holds, i.e. the supperradiant modes do not contribute to the atmosphere's
free energy. Hence,
\begin{eqnarray*}
\Gamma(E) &=& \frac{1}{\pi h}\frac{r_+^6}{\hh ^2}\times\\
           && \times\left[
            \int_0^E(E'-e\Phi )^2dE'\theta (-eQ)+
            \int_{e\Phi}^E(E'-e\Phi )^2dE'\theta (eQ)\right]\\
       &=&  \frac{1}{3\pi h}\frac{r_+^6}{\hh ^2}\times\\
           &&\times\left[
           \{(E-e\Phi )^3+(e\Phi)^3\}\theta (-eQ) +
            \{(E-e\Phi )^3\}\theta (eQ)\right] \\
       &\equiv& \Gamma(E)_-+\Gamma(E)_+
\end{eqnarray*}
and also
\begin{eqnarray*}
F &=& \left[
      T_H \Gamma (E)\ln(1-e^{(e\Phi -E )\beta})| _{E=0}^{\infty} -
      \int_0^{\infty}dE\frac{\Gamma (E)}{e^{(E-e\Phi )\beta}-1}
      \right]\theta (-eQ)\\
  &+& \left[
      T_H\Gamma (E)\ln(1-e^{(e\Phi -E )\beta})| _{E=e\Phi}^{\infty}-
      \int_{e\Phi}^{\infty}dE\frac{\Gamma (E)_-}{e^{(E-e\Phi )\beta}-1}
      \right] \theta (eQ)\\
  &=& -\int_0^{\infty}dE\frac{\Gamma (E)}{e^{(E-e\Phi )\beta}-1}
       \theta (-eQ)
      -\int_{e\Phi}^{\infty}dE\frac{\Gamma (E)_+}{e^{(E-e\Phi )\beta}-1}
       \theta (eQ)\\
  &\equiv& F_- +F_+,
\end{eqnarray*}
where we have assigned the  subscript $+$ to the  contribution of
particles
with the same sign of charge as the black hole's one, and  the subscript
$-$ to the contribution of their oppositely charged antiparticles. The
step-function $\theta (x)$ is  defined as usual: $\theta (x)=1$, if $x\ge
0$, and $\theta (x)=0$, if $x\le 0$. Substitution of $\Gamma$  yields:
\begin{eqnarray}
 F_-&=&-\frac{r_+^6}{3\pi h\hh ^2}
      \int_0^{\infty}dE\frac{(E-e\Phi )^3+(e\Phi )^3}
                  {e^{(E-e\Phi )\beta}-1}, 
      \;\; eQ\le 0;\\
F_+ &=& -\frac{r_+^6}{3\pi h\hh ^2}
      \int_{e\Phi}^{\infty}dE\frac{(E-e\Phi )^3}{e^{(e\Phi -E)\beta}-1}, 
      \;\; eQ\ge 0.
\end{eqnarray}
As long as we assume that $M\gg 1$, condition $\beta\gg 1$
holds for the black hole of any charge. Then  neglecting $-1$
in the integrand denominator and changing the variable of integration    
$E-e\Phi =y$ one gets:
$$
F_- = -\frac{r_+^6}{3\pi h\hh ^2}\left( T_He^{e\beta\Phi}(e\Phi )^3 +
      \int_{-e\Phi}^{\infty}y^3e^{-y\beta}dy\right) 
$$
\begin{equation}\label{free-}
\;\; =-\frac{r_+^6}{3\pi h\hh ^2}e^{e\beta\Phi}T_H^4
\left( 3(e\beta\Phi )^2-6e\beta\Phi +6\right),\;\; eQ\le 0;
\end{equation}
\begin{equation}\label{free+}
F_+ = -\frac{r_+^6}{3\pi h\hh ^2}
      \int_0^{\infty}y^3e^{-y\beta}dy
=-\frac{r_+^6}{3\pi h\hh ^2}6T_H^4,\;\;  eQ\ge 0.
\end{equation}
From eqs.(\ref{free-},\ref{free+}) we deduce that the contribution of
particles carrying a charge with sign opposite to the black hole's one is
negligible compared to the contribution of their antiparticles. This is
because they are pulled into the black hole by the electrostatic field as
soon as they  emerge and thus spend only a short time outside the horizon.
Conversely,  particles with the same sign of charge as the black
hole's one are pushed out of the horizon, then scatter off the effective
potential  $V_l$ (see eq.(\ref{V})) and then the lowest angular momentum
ones  escape to infinity, and the others return to the atmosphere. 

Note that $F_+$ does not depend on $e$. Thus, the thermal properties of
the charged atmosphere (and, hence, contribution of the horizon to these
properties) are the same as the neutral one and defined by the black hole
mass and charge completely.

In the limit of the neutral scalar field ($e=0$)
eqs.~(\ref{free+},\ref{free-}) reduce to
\begin{equation}\label{free0}
F_n= F_-(e=0)=F_+ =-\frac{2r_+^6}{\pi h\hh ^2}T_H^4.
\end{equation}
The  entropy of the neutral atmosphere is given by:
\begin{equation}\label{entropy0}
S_n=-\frac{\partial F_n}{\partial T_H}=\frac{8\hh }
{(4\pi )^3\pi h}.
\end{equation}
This coincides with the Hawking-Bekenstein entropy $S=\frac{1}{4}{\A}$
\cite{Hawking,Bekenstein} for the following value of the cutoff:
\begin{equation}
h=h_n=\frac{8\hh }{(4\pi )^3\pi ^2r_+^2}.
\end{equation}
The fact that $h$ depends upon $M$ and $Q$ is merely due to the
special choice of coordinates. Define the invariant distance as follows:
\begin{equation}\label{rho}
\rho=\int_{r_+}^{r=r_++h}ds=\int_{r_+}^{r=r_++h}\sqrt{-g_{rr}}dr\approx
\frac{2r_+}{\hh ^{1/2}}\sqrt{h}.
\end{equation}
So, $\rho _n\equiv\rho (h_n)=\pi ^{-5/2}2^{-1/2}$. 

We see that $\rho$ is a property of the horizon  independent of $M$ and
$Q$. The same result for the Schwarzschild black hole was obtained by 
't Hooft\cite{brick}.

In terms of the invariant distance the neutral atmosphere entropy reads:
$$
S_n=\frac{32}{(4\pi )^3\pi}\cdot\frac{r_+^2}{\rho _n ^2}.
$$

Suppose now that $e\not =0$. In any realistic black hole the following
inequality holds: $|eQ|\beta\gg 1$. Indeed, 
$$ 
|eQ|\beta\sim |eQ|\frac{M^2}{\hh}\sim |eQ|\frac{M}{\sqrt{1-(Q/M)^2}}\gg 1;
$$
which is clearly true since $Q\gg e_{electron}\sim 10^{-1}$ and $M\gg 1$.
Hence,  in the leading order  eq.~(\ref{free-}) gives:
$$
F_-=-\frac{r_+^6}{\pi h\hh ^2} T_H^2e^{e\beta\Phi}(e\Phi )^2. 
$$
Since $F_+$ does not depend on $e$ the entropy of the charged
atmosphere $S_c$ has only an exponentially small dependence on $e$:
\begin{eqnarray}
S_c &\equiv& S_++S_-  \nonumber\\
    &\approx&  \frac{32}{(4\pi )^4\pi}\cdot\frac{r_+^2}{\rho ^2}
    + \frac{r_+^6}{4\pi h}\frac{|e\Phi |^3}{\hh ^2}e^{-|e\Phi |\beta}
    \approx S_n.
\end{eqnarray}
It is seen  that at $\rho _c =\rho _n$ this expression returns to the
Hawking-Bekenstein formulae, as it has to, because, as we see, the thermal
properties of the atmosphere, which were used to define the cutoff, do not
depend on the scalar field charge $e$. It will be shown in the next
section that the invariant distance in the Kerr metric is also independent
of the atmosphere particle's angular momentum projection on the symmetry
axis $m$. 

For non-extreme black holes of mass greater than  $10^{15}$g it is
a Klein-paradox process that dominates the charged pair production, and
the emission rate is
governed by a Schwinger-type formula (for pair production in a constant
electric field). For black holes of smaller mass Hawking thermal process
dominates\cite{deWitt}.

\subsection{Horizon Fluctuations}\label{sec:fluctuations}
It was suggested in \cite{sanny} that due to the atmosphere of high
angular momentum particles, strong gravitational interactions
occur near the  Schwarzschild black hole horizon at an invariant distance
of the order of $M^{1/3}$. This is because the total energy of the black
hole's atmosphere fluctuates as any thermodynamical variable.  
Let us extend this idea to the case of the charged black holes.

According to the rules of thermodynamics, given the energy of the 
atmosphere  $U(\rho )$ between the surfaces $r=r_++h(\rho)$ and $r=r_1$, and the
number of particles $N(\rho)$ between these surfaces, the black
hole's mass fluctuation  between $r=0$ and $r=r_++h(\rho)$  is 
\begin{equation}\label{delta}
\Delta M(\rho )=\Delta U(\rho )\sim\frac{U(\rho )}{\sqrt{N(\rho )}}, 
\end{equation}
provided that Riemann tensor is sufficiently small near the horizon
 ${\bf R}\sim M^{-2}\ll 1$, so that we can apply the usual flat spacetime 
thermodynamic rules in the vicinity of the horizon.  
We have used the fact that the total energy $M$
 defined in (\ref{metricC}) is fixed from the point of view of an external
observer. The  energy of the atmosphere above the surface
$r=r_++h(\rho)$ is given by:
$$
U(\rho)=\frac{\partial}{\partial\beta}(\beta F)\sim\frac{\hh}{\rho ^2}.
$$
The number of  particles is $N\sim S$. Therefore:
$$
\Delta M(\rho)\sim\frac{\hh}{r_+\rho}.
$$

Right now we are facing problem mentioned in the
section~\ref{sec:introduction}: we had to use renormalized total energy
$U^{ren}$ instead of $U$. However, as
far as only the variation of the total energy is needed, one may replace   
$\Delta U^{ren}$ by $\Delta U$. Indeed, all divergent terms in the
energy-momentum tensor are functions of the  curvature tensor components
and their derivatives only\cite{Birrel}, so they are canceled from the
expression for fluctuation of the total matter energy\footnote{In 
the ref.~\cite{Israel}  S.Mukohyama and W.Israel argued 
that if one correctly identifies the ground state --- the Boulware vacuum,
then the total energy near the black hole horizon is finite. They use this
fact to advocate the brick-wall model. Conclusions of this section are
independent  of choice of the ground state; the thermal fluctuations
depend only on the number of modes near the horizon.}.

Let us estimate how the fluctuating mass gives rise to uncertainty in the
location of the horizon. A point $r'$ is outside the horizon if 
$ r'-r_+(M(r'))>0$,
where $M(r')$ is the black hole energy between $r=0$ and $r=r'$.
Clearly, if 
\begin{equation}\label{in-out}
\Delta (r'-r_+(M(r')))=\Delta (r_+(M(r'))>\ r'-r_+(M(r'))=h,
\end{equation}
 then the point $r'$ is in the superposition of being inside and outside
the horizon. Using eq.~(\ref{hor}) we obtain:
\begin{equation}\label{Deltar+}
\Delta r_+=\frac{2\Delta Mr_+}{r_+-r_-}=\frac{1}{\rho}.
\end{equation}
From eq.~(\ref{rho}) we see that relation $\Delta r_+<h$ holds if
$$
\frac{1}{\rho}<\frac{\rho ^2\hh}{r_+^2};
$$
that is, the semi-classical approach is valid as long 
as
\begin{equation}\label{rhomin}
\rho >\rho _{min}=\left( \frac{r_+^2}{r_+-r_-}\right)^{1/3}.
\end {equation}
This equation  implies that in Schwarzschild metric 
$\rho _{min}= M^{-1/3}$, see ref.~\cite{sanny}.   In the case of extreme
black hole $\rho _{min}$  becomes infinite. 


We have derived eq.~(\ref{Deltar+}) neglecting fluctuations of the
atmosphere's total charge ${\cal Q}$.   Nevertheless, it  remains valid if
take account of the ${\cal Q}$-fluctuation. Indeed,
\begin{eqnarray*}
\Delta r_+ &=& \sqrt{\left(\frac{r_+}{r_+-r_-}\right) ^2\Delta M^2
           +\frac{Q^2}{\hh ^2}\Delta Q^2}\\
          &=& \frac{r_+}{r_+-r_-}\sqrt{\Delta U^2 
         +\frac{Q^2}{r_+^2}\Delta {\cal Q}^2}=
        \frac{r_+}{r_+-r_-}\Delta\left( U-\frac{Q}{r_+^2}{\cal Q}\right).
\end{eqnarray*}             
Since the quantity in the curly brackets of the last equation is
additive we estimate it's fluctuation  as follows:
$$
\Delta\left( U-\frac{Q}{r_+^2}{\cal Q}\right)\simeq
\langle E-\frac{Q}{r_+^2}e\rangle \sqrt{N}\simeq T_H\frac{r_+}{\rho}.
$$
Putting all this together leads to eq.~(\ref{Deltar+}).


\subsection{ Kerr Black Holes}\label{sec:Kerr}

Consider a  rotating
black hole with projection of angular momentum on the
symmetry axis $J$ (called simply black hole's "angular momentum"). The
metric in Boyer-Lindquist coordinates is given by\cite{Landau}:
$$
ds^2 = \left(1-\frac{2Mr}{\Sigma ^2}\right) dt^2
         -\frac{\Sigma ^2}{D}dr^2-\Sigma ^2d\theta ^2 
$$
\begin{equation}\label{Kerr-Newman}
     -\left(r^2+a^2+\frac{2Mra^2}{\Sigma ^2}\sin ^2\theta\right)
        \sin ^2\theta d\f ^2
        +\frac{2\cdot 2Mra}{\Sigma ^2}\sin ^2\theta d\f dt.
\end{equation}
where $a=J/M$, 
\begin{equation}
D =r^2-2Mr+a^2,\;\;\; \Sigma ^2 =r^2+a^2\cos ^2\theta.
\end{equation}

 The event horizon $r_+$ and surface $r_-$ are defined as
\begin{equation}\label{horkerr}
r_{^+_-}=M^+_-\sqrt{M^2-a^2}.
\end{equation} 
There is a surface such that no static observer can exists inside it
(frame dragging effect). Definition of this surface, called static limit,
is
$$
r_0(\theta )=M+\sqrt{M^2-a^2\cos ^2\theta}.
$$

As in the case of the charged black hole there exists Hawking radiation
with the temperature $T_H$ given by \cite{deWitt}:
$$
T_H=\frac{\hh}{{\A}}=\frac{\hh}{4\pi(r_+^2+a^2)}=\frac{\hh}{8\pi Mr_+}.
$$
Inertial observer at infinity measures flux of energy and angular
momentum from the rotating black hole. We are interested to study what
happens to the system of Hawking particles reflected back to the black
hole by the space-time curvature. 

Recall that in the Kerr metric one can separate variables in the wave
equation (\ref{waveequation})\cite{deWitt,Unruh}. The basis functions are:
$$
u_{E\lambda_{lm}m}(x)=(r^2+a^2)^{-1/2}
R_{lm}(r)S_{lm}(\cos\theta )
e^{im\f}e^{-iEt},
$$
where $S_{lm}$ is the spherical harmonics with eigenvalue $\lambda
_{lm}(aE)$, $l=0,1,2\ldots$, $m=-l,-l+1,\ldots l-1,l$.   

The radial part $R_{lm}(r)$ satisfies the following equation:
$$
[\frac{d^2}{dr^{*2}}-V_{Elm}(r)]R_{lm}(r)=0,
$$
where $r^*$ and $V_{Elm}(r)$ are defined by the following relations:
\begin{equation}\label{rstar}
\frac{dr^*}{dr}=\frac{r^2+a^2}{D},
\end{equation}
\begin{eqnarray}
V_{Elm}(r) &=& -(E-m\frac{a}{r^2+a^2})^2+
                   \lambda _{lm} (aE)\frac{D}{(r^2+a^2)^2}+ \nonumber\\
                   & & \frac{2(Mr-a^2)D}{(r^2+a^2)^3}+
                     \frac{3a^2D^2}{(r^2+a^2)^4}.
\end{eqnarray}   

 The centrifugal barrier is 
$$ 
V_{Elm}^{c}(r)=\lambda _{lm} (aE)\frac{D}{(r^2+a^2)^2}.
$$
It is attractive for $r_+\le r<r_1$ and repulsive for $r>r_1$, where 
$r_1$ is a physically acceptable root of equation $\frac{dV^c}{dr}=0$:
\begin{equation}
r_1=M+2\sqrt{M^2-\frac{1}{3}a^2}\cdot\cos{\frac{\alpha}{3}},
\end{equation}
where
$$ 
\cos\alpha =\frac{1-a^2/M^2}{(1-a^2/3M^2)^{3/2}}. 
$$
Once again the tunneling through the angular momentum barrier may be
neglected for all but the lowest angular momentum modes since 
$V_{lm}^{c}\sim\lambda _{lm} \sim l^2$. 

The thermal atmosphere of the black hole extends from the horizon up to
the surface  $r=r_1$, where $r_1$ is the turning point of the centrifugal 
barrier.  Study of two limit cases $a=0$ and $a=M$ shows that $r_1>r_0$
for any  black hole. In other words the ergosphere lies under the external
boundary of the atmosphere. 

Let us calculate a number of wave equation solutions in the vicinity of
the black hole. This number is equal to the twice of number of quantum
states of the atmosphere, because of two-fold degeneration of energy
levels (along the symmetry axis). In the WKB approximation:
$$
\Gamma  = 2(2\pi)^{-3}\sum_m\int d\varphi\int dr\int dp_r\int
d\theta\int dp_{\theta}.
$$
Here the covariant radial momentum $p_r$ associated with the differential
operator $\hat{p}_r$ in the wave equation is given by \cite{Landau}:
\begin{equation}\label{pr}
p_r^2=D^{-2}[(r^2+a^2)E-am]^2-\lambda_{lm} D^{-1},
\end{equation}
and the polar momentum $p_{\theta}$ associated with $\hat{p}_{\theta}$:
\begin{equation}\label{ptheta}
p_{\theta}^2=\lambda_{lm} -(aE\sin\theta-\frac{m}{\sin\theta})^2.
\end{equation}

Introduce a  cut-offs $h$ and $L$ which are  defined as
$\phi (x)=0$ if $r\ge L$ and $r\le r_++h(\theta )$. Now allow $h$ to be
some function of $\theta$.  


Let us change variables of integration replacing the  pair
$(p_r,p_{\theta})$ by the pair of constants of motion $(\lambda ,E)$
(here $\lambda _{lm}\equiv \lambda$ ). Jacobian  of this transformation is
\begin{eqnarray*}
&&\frac{\partial (p_r,p_{\theta})}{\partial (\lambda ,E)}=\\
&&\frac{2D^{-2}(r_+^2+a^2)^2(E-m\Omega _H)-
     2D^{-1}(aE\sin\theta -m/\sin\theta)a\sin\theta}
     {4p_r(\lambda ,E)p_{\theta}(\lambda ,E)}
\end{eqnarray*}
Here $\Omega _H=a/2Mr_+$ is the angular velocity of the black
hole\cite{Misner}.
Since the main contribution to the atmosphere's partition function arises
from the region near the horizon ($r\approx r_++h$) we neglect the second
term in the last equation  and obtain (note eq.~(\ref{D})) 
\begin{eqnarray*} 
\Gamma (E)&=& 2(2\pi )^{-3}\sum_m\int d\f\int d\theta\int dr\int d\lambda 
\int dE \frac{\partial (p_r,p_{\theta})}{\partial (\lambda ,E)}\\
&\approx& 2(2\pi )^{-3}2\pi h\sum_m\int d\theta\int d\lambda\int dE\times\\
  &\times&
  \frac{D^{-2}(r_+^2+a^2)^2(E-m\Omega _H)+O(h^{-1})}
       {2\sqrt{2D^{-2}(r_+^2+a^2)^2(E-m\Omega _H)^2-\lambda D^{-1}}
         \sqrt{\lambda-(aE\sin\theta -m/\sin\theta )^2}},
\end{eqnarray*} 
where $D$ near the horizon is given by (\ref{D}).

Now, the free energy for large $\beta$ is
\begin{eqnarray*}
F &=& T_H\sum_m \int d\Gamma (E,m)\ln\left( 1-e^{(m\Omega _H-E)\beta}
      \right) \\
 &\approx& -\frac{h}{(2\pi )^2}\frac{T_H(r_+^2+a^2)^2}{D}
        \int d\theta\int_0^{\lambda _{m}} d\lambda\int_0^{\infty}dx
        \frac{xe^{-x\beta}}{\sqrt{(r_+^2+a^2)^2x^2-\lambda D}}\times\\
  &&\times
        \int dm\frac{1}{\sqrt{\lambda -[a\sin\theta (x+m\Omega _H)-
            m/\sin\theta ]^2}}+O(e^{-|m|\Omega_H\beta})\\
   &=& -\frac{1}{2\pi}T_H(r_+^2+a^2)^3\int d\theta\sin\theta
          \int_0^{\infty}dxx^2e^{-x\beta}D^{-2}\\
   &=& -\frac{1}{\pi}T_H^4\frac{(r_+^2+a^2)^3}{\hh ^2}\int
              d\theta\frac{\sin\theta}{h}
\end{eqnarray*}
We have introduced a new variable of integration $x=E-m\Omega _H$ and
taken account of the fact that contribution of particles with $m<0$ is of
the order of $O(e^{-|m|\Omega _H\beta})$ (like in the charged black hole
case). The radial momentum $p_r$ vanishes at $\lambda =\lambda _{m}$. 

It is easy to convince, that the entropy is given by
\begin{equation}\label{entropyKerr}
S=\frac{4}{\pi}T_H^3\frac{(r_+^2+a^2)^3}{\hh
                  ^2}\int_0^{\pi} d\theta\frac{\sin\theta}{h}.
\end{equation}
This formula is consistent with  eq.~(\ref{entropy0}). (To see this put  
 $Q=0$ in eq.~(\ref{entropy0}) and $a=0$ in eq.~(\ref{entropyKerr})and
compare).


Let us define invariant distance $\rho$ as in eq.~(\ref{rho}):
\begin{equation}\label{rhokerr}
\rho =\sqrt{\frac{r_+^2+a^2\cos ^2\theta}{r_+-r_-}}\sqrt{h(\theta )}.
\end{equation}
Substitution to (\ref{entropyKerr}) gives
$$
S=\frac{4}{\pi(4\pi )^3}\hh\int_0^{\pi}\frac{d\theta\sin\theta}{h}=
\frac{4}{\pi(4\pi )^3}\int_{-1}^{1}\frac{d\xi(r_+^2+a^2\xi ^2)}{\rho^2(\xi
)}.
$$ 
We may choose $h(\theta )$ in such a way that  this equation will coincide
with the  Hawking- Bekenstein expression for the black-hole entropy.
For example, take $h(M,a|\theta )=\bar{h}(M,a)/(\cos^{-2}\theta +3).$
The problem is that there are infinitely many ways to do this, and thus,
one
cannot fix the $\theta$ dependence of $h$. Despite this difficulty,
however,  we see that whatever function $h(\theta )$ that matches the
Hawking-Bekenstein expression one chooses, the invariant cutoff  $\rho$
does not depend on the black hole parameters $M$ and $J$ and on the
atmosphere's thermal properties. It is characterized by the event
horizon only.  

The fact that we do not know the exact form of the function $h(\theta )$ 
does not matter, if we are interested only in ratios of such
thermodynamical variables as the total  energy, entropy etc. In this
case we need not know  this function. This is the case if we are going to
calculate the horizon fluctuations. We repeat the by now familiar procedure
of  sec.~\ref{sec:fluctuations} and arrive at the formula~(\ref{rhomin})
with $r_+$ and $r_-$ defined by eq.~(\ref{horkerr}).
Note that deriving this result we have effectively averaged 
thermodynamical variables over all possible choices of $h$ which yield
the Bekenstein-Hawking entropy. 

It is straightforward to generalize this result to the Kerr-Newman black
hole. We will refer to eq.~(\ref{rhomin}) as such generalization with 
\begin{equation}
r_{^+_-}=M^+_-\sqrt{M^2-a^2-Q^2}.
\end{equation} 
 
It should be emphasized, that despite the fact that the $\rho _{min}$
becomes infinite for the extreme black holes, the condition $h\ll M$ under
which all calculations were done still  holds. Indeed,  using
eq.~(\ref{rhokerr}) this condition reads
\begin{equation}\label{cond}
\rho\ll \frac{M^{3/2}}{(M^2-a^2-Q^2)^{1/4}}.
\end{equation}
It is seen by substituting $\rho_{min}$ from eq.~(\ref{rhomin}) into
eq.~(\ref{cond}) that the $\rho _{min}$ always matches (\ref{cond}).

\section{ Shock Wave Model}\label{sec:shock}

In this section we shall show that strong gravitational interaction
in the vicinity of the charged  black hole occurs at $\rho$
given by
eq.~(\ref{rhomin}) by studying the gravitational interactions between an
in-falling particle and the  thermal atmosphere. 

It is shown in the Appendix, that propagation of a massless particle on 
the charged black hole background can be described by the shock wave of 
a special form, so that the resulting metric is given by 
eqs.~(\ref{sh.w.}),(\ref{A}),(\ref{g}) and (\ref{fff}). Consider some
massless particle falling into  the
black hole. The atmosphere (Hawking) particles interact with the shock
wave generated by this particle. What is the probability that the state of
the atmosphere is the same after the interaction took place?

At first, we shall study interaction of the in-falling particle with
the one test atmosphere particle. Note that the  Reissner-Nordstr\o m
geometry  (\ref{metricC}) takes the same approximate form as the
Schwarzschild one near the horizon:
$$
ds^2=\frac{\rho ^2\hh ^2}{4r_+^4}dt^2-d\rho ^2-dx^2-dy^2,
$$
provided that we study a region with transverse distances much smaller
than $M$. Here $\rho$ is defined in (\ref{rho}) independently of the 
black hole charge. Denote
$$
\M\equiv\frac{r_+^2}{2\hh},
$$
then this metric reads
\begin{equation}\label{metricM}
ds^2=\frac{\rho ^2}{(4\M )^2}dt^2-d\rho ^2-dx^2-dy^2.
\end{equation}
Parameter $\M$ plays for  metric of the charged black hole the same
role as the mass $M$ for the neutral one.    

We now follow the arguments of ref.~\cite{sanny} for the Schwarzschild 
black hole.
Define Rindler coordinates 
\begin{equation}\label{Rindler}
u=T+z=\rho e^{t/4\M},\;\;
v=T-z=-\rho e^{-t/4\M}.
\end{equation}
Then metric (\ref{metricM}) is simply Minkowski space in these
coordinates
\begin{equation}\label{Minkowskimetric}
ds^2=dudv-dx^2-dy^2.
\end{equation}
Let $k_{\mu}$,  $p_{\mu}$  be the  momenta of in-going and atmosphere
(Hawking) particles respectively.
The gravitational field of massless point-like particle in Minkowski
space is described by the line element \cite{Hooft}
\begin{equation}\label{metricH}
ds^2=du\left[ dv+2k^v\ln\left(\frac{\tilde{x} ^2}{M^2}\right)
           \delta (u-u_0)du\right] -dx^2-dy^2,
\end{equation}
where $\tilde{x} ^2=x^2+y^2$, $u=T+z$ and $v=T-z$. The massless particle
moves in the $v$ direction with constant $u_0$ and momentum $k^v$.
  
The effect of the shock wave on the massless particle  propagating in the
metric (\ref{metricH}) with initial momentum $p_{\mu}$ is a discontinuity
in the $v$ direction at $u=u_0$:
\begin{equation}\label{shift}
\Delta v=-2k^v\ln\left(\frac{\tilde{x} ^2}{M^2}\right),
\end{equation} 
and a refraction in the transverse direction:
\begin{equation}\label{refraction}
p_x(u)-p_x=\frac{4k^v}{\tilde{x}^2}xp_v\theta (u-u_0), 
\end{equation}
and similarly for $p_y(u)$.

By forming  wave packets describing a high angular momentum Hawking
particle before and after crossing the shock wave, and then calculating
their scalar product one finds that the probability to be in the same
state after crossing the shock wave is\cite{sanny}  
$$
P_1 \sim 1-\frac{\M ^2\epsilon^2}{\rho^4},
$$
where $\epsilon$ is the energy of the atmosphere particle. Also,
the probability for one particle in the atmosphere to have changed angular
momentum is 
$$
P_{\Delta l\neq 0}=\frac{\M ^2\epsilon ^2}{\rho^4}.
$$  
This is the result of interaction of in-falling particle with one
atmosphere particle.

The number of particles which are affected by the shock wave of the
in-going particle when it reaches $\rho$ can be deduced from 
eqs.~(\ref{entropy0},\ref{rho}):
$$
N(\rho )\sim S(\rho )\sim\frac{r_+^2}{\rho ^2}.
$$
Thus the probability for the atmosphere above $\rho$ to remain in the
initial state is
$$
P_{tot}=P_1^{N(\rho )}=
         \left(1-\frac{\M ^2\epsilon ^2}{\rho ^4}\right)^{N(\rho )}
\sim e^{-\M ^2\epsilon ^2r_+^2/\rho ^6}.
$$
We see that the state of the atmosphere is changed when the particle
reaches $\rho=(\M r_+\epsilon )^{1/3}$. 

The minimal $\epsilon$ one can consider is $\frac{1}{r_+}$ since
otherwise the wavelength of the in-going particle is larger than the
radius of the black hole. Thus the minimal $\rho$ is given by the
following eq.
$$
\rho _{min}=\M^{1/3}\sim\left( \frac{r_+^2}{r_+-r_-}\right)^{1/3}.
$$
which coincides with eq.~(\ref{rhomin}).

It can be easily shown using the same arguments as in ref.~\cite{sanny}
that the information carried by an in-going massless spin-less charged
particle
is encoded in the state of the atmosphere when the particle reaches $\rho
_{min}$.

\section{Discussion}\label{sec:discussion}

We saw in the previous sections that the high angular momentum Hawking
particles get reflected by the centrifugal barrier near the black hole
horizon. It turns out that the number, total energy and other statistical
quantities of these particles,  which are  found in the
region over the horizon, depend only on the black hole
parameters. This allows to speak about the atmosphere of Hawking
particles near the horizon, as about pure quantum geometrical phenomenon. 

Since the Hawking radiation has a Planckian structure it is natural to
say that a system consisting of a black hole and its radiation is in the
thermal equilibrium state. One is interested to know  thermodynamical
parameters of the atmosphere in this state. In order to calculate entropy 
and the free energy of the atmosphere we counted the number of modes of
the scalar field near the horizon, and then used well-known statistical
thermodynamic formulae. Both entropy and the free energy are proved to be
divergent at the horizon.

In order to deal with these divergences we introduced a cutoff at the  surface
$r=r_++h$. The leading order term $O(h^{-1})$ in the expression for the
entropy is the contribution of the horizon to the total atmosphere entropy.  
The value of $h$ was fixed by the requirement  that the
black hole entropy  be a quarter of the horizon area, according to
the Hawking-Bekenstein formula. This program was carried out for the
Schwarzschild black hole in ref.~\cite{brick}, and for charged and
rotating black holes in sections~\ref{sec:charge} and \ref{sec:Kerr}
respectively. We found that the value of the
cutoff expressed in terms of the invariant distance is the same for all
black holes independently of their mass, charge and angular momentum and
is of order unity. At this (Planck) scale the semi-classical approach 
breaks down.

Up to now the back reaction of the emerging radiation was neglected.
Accounting for the back reaction gives rise to the black hole's mass
decreasing and to the black hole's parameters fluctuation. The first
phenomenon can be neglected at short time scales, provided that the black
hole's energy is much larger than the Planck mass. The second one is
responsible for the strong gravitational interactions occurring near the
horizon.

Indeed, by studying the atmosphere's mass fluctuations and interactions
between Hawking and in-coming particles, we showed that the classical
trajectories near the horizon cease to exist at invariant distances
smaller than $\rho_{min}=[r_+^2/(r_+-r_-)]^{1/3}.$ This means that one
cannot continue to use a non-quantized background  metric at this scale,
the semi-classical approach becomes invalid.
     
If the black hole has small charge and angular momentum then $\rho _{min}
\ll M$  and the question is: how does the high frequency
part of the  Hawking radiation spectrum  affected by the gravitational
interactions. Our analysis does not provide the analytical answer to this
 question. However, the fact that  information is encoded out
of the horizon, may be the reason {\em pro} the possibility of 
$S$-matrix construction advocated by 't Hooft.

An essentially new effect arises, when we turn our attention to the
 extreme black holes. As the distance between horizon $r_+$ and the
surface $r_-$ becomes smaller, the strong gravitational interactions
occur at greater invariant distance, the smaller frequency Hawking
quanta are affected, and so the whole spectrum becomes non-thermal. This
in turn implies that in the case of the extreme black hole the information
of an in-going particle is encoded at distances much larger than black
hole's radius. In this case, the semi-classical approach  breaks
down at all scales.

Despite the fact that the minimum invariant distance from the
horizon $\rho_{min}$ at which the semi-classical theory still holds
becomes infinite for the extreme black holes, its counterpart in the
Boyer-Lindquist coordinates $h_{min}\equiv h(\rho_{min})$ vanishes.
However it still remains greater than value of the cutoff. Indeed,   
definition of the invariant distance implies that in general
$$
h_{min}\sim\rho_{min}^{-1}\sim\frac{\hh ^{1/3}}{M^{2/3}}.
$$
The cutoff in these coordinates is
$$
h_{{\rm cutoff}}\sim \frac{\hh}{M^2}
$$
So, $h_{{\rm cutoff}}<h_{min}$ for all black holes.  
Thus, one cannot reach the extreme black hole horizon (naked singularity)
without entering a region below the $h_{min}$ which cannot be
investigated with the semi-classical theory. We
believe that the gravitational interactions which occur in this region
somehow prohibit the particle to reach the horizon (and to violate 
causality).

There is another simple argument which explains why  one cannot reach the
extreme black hole horizon in spite the fact that the non-invariant cutoff
vanishes. The proper time it takes  a freely falling observer to reach
the horizon from some point $R$ out of the horizon is infinite  because
this time is just the invariant distance
$$
\Delta\tau =\rho\sim\frac{M}{r_+-r_-}\sqrt{2M-R},   
$$
which is infinite for all $R$ in our case.

As we explained, the semi-classical approach is wrong if the radial
coordinate in the Boyer-Lindquist coordinates is smaller than
$r_++h_{min}$. This leads to some inconsistency when we use the brick
wall model, since we count semi-classical modes  between $r_++h_{{\rm
cutoff}}$ and $r_++h_{min}$ also. However, we can neglect the contribution
of this region, since the number of modes in the atmosphere is 
determined by the phase space volume in the nearest vicinity of the
cutoff. $\Gamma(\rho_{min})$ is of order $O((M\hh)^{2/3})$
which is negligible
compared to the leading order term $O(M^2)$.  Moreover, the explicit
(quantum gravitational) accounting for the processes occurring in this
region should  not change the estimate of $\rho_{min}$, since a
completely different approach,--- shock wave model, which does not use the
counting of states, gives the same result. It is interesting to note that
in the extreme black hole case the problematic region shrinks and has
zero measure.
   
We did not discuss so far  how an external observer learns about the
black hole's horizon fluctuations. We noted, that the black hole mass is
fixed in his frame (apart of small decrease caused by the back reaction).  
So, in order to obtain information about the horizon fluctuations, the
external observer should study trajectories of test particles which pass
near the horizon. The fluctuating horizon may trap, with some
probability, a particle with definite quantum numbers. And it may release
another particle, which, in general, will not bear  the same quantum
numbers, since we have no information about the state of order or
disorder of  matter inside the black hole. Therefore, the out-going
geodesics  which pass near the horizon will be thermally averaged and
will differ from the classical geodesics.

The external  observer may also ask another observer who is found near the
horizon, to send him information about the atmosphere's state. Then he
must process this information bearing in mind that the Hawking temperature
of the observer in the atmosphere is blue-shifted, and the light
signals are red-shifted compared to those in his frame. 
It may be interesting to investigate this issue in more detail.  

The fact that the Hawking radiation is merely fluctuations of vacuum, 
may  put in doubt our approach to the atmosphere as a system of real
on-shell particles. As was pointed out in ref.~\cite{sanny}, it
is expected to be a valid approximation if the $S$-matrix ansatz of 't
Hooft is correct.

In summary, our analysis suggests that  strong gravitational interactions
occur near the black hole horizon at an invariant distance of  order 
$\rho_{min}=[r_+^2/(r_+-r_-)]^{1/3}$. At smaller distances the
semi-classical approach breaks down. 


\appendix
\section{Appendix. Shock wave on charge black hole background}

In the Appendix we shall show that the shock wave of a special form 
(\ref{sh.w.}), generated by a point-like massless particle  can propagate
on the charged non-extreme black hole background.

Recall the general result obtained by Dray and 't Hooft\cite{Hooft}:
Given a solution of the vacuum Einstein equations of the form
\begin{equation}\label{theorem}
ds^2=2A(u,v)dudv+g(u,v)h_{ij}(x^i)dx^idx^j;
\end{equation}
if the following conditions hold
\begin{equation}\label{conditions}
A_{,v}|_{u=0}=0=g_{,v}|_{u=0},
\end{equation}
\begin{equation}\label{func.f}
\frac{A}{g}\Delta f-\frac{g_{,uv}}{g}f=32\pi pA^2\delta (\tilde{x}),
\end{equation}
where $f=f(x^i)$ represents the shift in $v$, $\Delta f$ is the Laplacian
of $f$ with respect to the 2-metric $h_{ij}$ and $\tilde{x}$ is the
transverse distance, then the shift in $v$ at $u=0$ can be introduced so
that the resulting space-time solves the field equation with a photon at
the origin $\tilde{x}=0$ of the $(x^i)$ 2-surface and $u=0$. The resulting
metric is then described as follows:
\begin{equation}\label{sh.w.}
d\hat{s}^2=2A(u,v+\theta f)du(dv+\theta f_{,i}dx^i)+g(u,v+\theta
f)h_{ij}dx^idx^j.
\end{equation}

Let us verify  whether conditions of this statement are satisfied in the 
case we are interested in. 
At first, we introduce new coordinates 
$\tilde{u},\tilde{v}$ which are labels for outgoing and in-going, radial,
null geodesics\cite{Misner}. The geodesics are solutions of the following
equation:
\begin{equation}\label{newcoord1}
ds^2=0=\frac{D}{r^2}dt^2-\frac{r^2}{D}dr^2.
\end{equation}
Hence they are given by:
\begin{equation}\label{tildes}
\tilde{u}=t-r^*,\;\;\; \tilde{v}=t+r^*,
\end{equation}
where $r^*$ is defined by (\ref{tort.}) and reads as follows 
\begin{equation}\label{r*}
r^*=r+M\ln\frac{D}{D_0}+\frac{2M^2-Q^2}{\hh}\ln\frac{r-r_+}{r-r_-},
\end{equation}
and $D_0$ is integration constant.

The line element (\ref{metricC}) in terms of new coordinates reads:
\begin{equation}\label{lineel1}
ds^2=\frac{D}{r^2}d\tilde{v}d\tilde{u}-r^2(d\theta ^2+\sin ^2\theta d\f
^2).
\end{equation}
It has obvious pathology at the horizon: $D(r)|_{r=r_+}=0$. In order to
remove it we change coordinates again\cite{Misner}. Using 
eqs~(\ref{tildes}),(\ref{r*}) one gets
\begin{equation}\label{usefrel}
e^{(\tilde{v}-\tilde{u})/(4M)}= e^{r^*/2M}=
e^{r/2M}\left(\frac{r-r_+}{2M}\right)^{\frac{1}{2}+\alpha}
                                   (r-r_-)^{\frac{1}{2}-\alpha},
\end{equation}
where
$$
\alpha=\frac{2M^2-Q^2}{2M\hh},
$$
and $D_0$ is fixed using requirement that  equation (\ref{usefrel})
must yield the correct result at $Q=0.$
Thus, new coordinates are defined as follows
\begin{equation}\label{newcoord2}
u=-e^{-\tilde{u}\beta /4M};\;\;\; v=e^{\tilde{v}\beta /4M},
\end{equation}
where $\beta$ is a certain constant to be fixed in such a way as to remove
the pathology in the metric (\ref{lineel1}).
We have
$$
\frac{D}{r^2}d\tilde{u}d\tilde{v}=
 \frac{16M^2(2M)^{(\frac{1}{2}+\alpha)\beta}}{\beta^2r^2}
       e^{-\frac{\beta r}{2M}}(r-r_+)^{1-(\frac{1}{2}+\alpha)\beta}
                   (r-r_-)^{1-(\frac{1}{2}-\alpha)\beta}dudv,
$$
where eqs.~(\ref{usefrel}) and (\ref{newcoord2}) were used.
Clearly, the value  of $\beta$ we need is
\begin{equation}\label{beta}
\beta=\beta_0\equiv\frac{1}{\frac{1}{2}+\alpha}.
\end{equation}
Thus, by substitution this into (\ref{lineel1}) the line element reads
\begin{equation}\label{lineel2}
ds^2=\frac{32M^3}{\beta_0^2r^2} e^{-\frac{\beta_0 r}{2M}}
               (r-r_-)^{1-(\frac{1}{2}-\alpha)\beta_0}
              dudv- r^2(d\theta ^2+\sin ^2\theta d\f^2).
\end{equation}
In coordinates $u,v,\theta,$ and $\f$ metric of the charged black hole
(\ref{lineel2}) has  no pathologies. The only singularity at $r=0$
is a physical one.

Functions $A$ and $g$ defined in eq.~(\ref{theorem}) are
\begin{equation}\label{A}
A=\frac{16M^3}{\beta_0^2r^2} e^{-\frac{\beta_0 r}{2M}}
(r-r_-)^{1-(\frac{1}{2}-\alpha)\beta_0},
\end{equation}
\begin{equation}\label{g}
g=-r^2.
\end{equation}
Note the useful relation following from eqs.~(\ref{usefrel})
and (\ref{newcoord2}):
$$
uv=- e^{\frac{\beta_0 r}{2M}}\frac{r-r_+}{2M}
             (r-r_-)^{(\frac{1}{2}-\alpha)\beta_0}.
$$
It shows that $A$ and $g$ are functions of the product $u\cdot v$, so
$$
A(u\cdot v)_{,v}\propto u;\;\;\; g(u\cdot v)_{,v}\propto u,
$$
and therefore conditions (\ref{conditions}) are satisfied.

Also, 
$$
g_{,u}=-\frac{4MD}{\beta_0 r} e^{-\frac{\beta_0 r}{2M}}\frac{2M}{r-r_+}
(r-r_-)^{(\alpha -\frac{1}{2})\beta_0}v,
$$
$$
g_{,uv}=\frac{A}{r}\left[2(r-M)-D(\frac{1}{r}+\frac{\beta_0}{2M})-(r-r_-)
+(r-r_+)(\alpha-\frac{1}{2})+\frac{\beta_0 r^2}{2M}\right].
$$
Following ref.~\cite{Hooft} we arrange the  coordinates so that the 
massless particle is at $\theta =0=u$. 
Then, upon substitution of these relations into equation (\ref{func.f})
one obtains
\begin{equation}\label{ff}
\Delta f-\lambda(M,Q)f=-2\pi\kappa(M,Q)\delta(\theta),
\end{equation}
where
$$
\lambda(M,Q)=\frac{1}{r_+}\left[2(r_+-M)-\hh +
                             \frac{\beta_0 r_+^2}{2M}\right],
$$
$$
\kappa(M,Q)=2^8p\frac{M^3}{\beta_0^2}\hh^{1+(\alpha-\frac{1}{2})\beta_0}
e^{-\frac{\beta_0 r}{2M}}.
$$
Note that $\lambda(M,0)=1$, and $\kappa(M,0)=2^9pe^{-1}M^4$. This is the
result obtained by Dray and 't Hooft.

 We can find solution to eq.~(\ref{ff}) by expanding $f$ in terms of
spherical harmonics. However, since only spherical harmonics with $m=0$
contribute, it is suffice to expand $f$ in terms of Legendre polynomials
$P_l(\cos\theta)$. Then
\begin{equation}\label{fff}
f=\kappa\sum_{l=0}^{\infty}\frac{l+\frac{1}{2}}{l(l+1)+\lambda}
                            P_l(\cos\theta).
\end{equation}
Since the asymptotic behavior of the Legendre polynomials at large $l$ is 
$$
P_l\sim \frac{1}{l^{1/2}},
$$
this sum  converges for all $\lambda$ such that $0<\lambda\le 1$, i.e.
for non-extreme black holes. $\lambda =0$ corresponds to the extreme black
holes $|Q|=M$; in this case $\kappa(M,M)=2^8pM^3$ and
$f$ diverges according to the following law:
$$
f^{{\rm ext}}=\frac{\kappa(M,M)}{2\lambda(M,M)}
\sim\frac{1}{\sqrt{M-|Q|}}.
$$


\section*{Acknowledgments}
The author would like to thank  N.Itzhaki, F.Englert and
especially A.Casher for  fruitful discussions on related topics. 

\vfill

%

\end{document}